\documentclass{aa}  
\usepackage{graphicx}
\usepackage{txfonts}
\usepackage{mhchem}
\usepackage{hyperref}
\makeatletter
\newcounter{chemeqn,xcolor}

\makeatother

\begin{document}

   \title{New SiS destruction and formation routes via neutral-neutral reactions and their fundamental role in interstellar clouds at low and high metallicity values}


   \author{
Edgar Mendoza,\inst{1}
Samuel F. M. Costa,\inst{2}
Miguel Carvajal, \inst{1,3} 
Sérgio Pilling, \inst{4}
Márcio O. Alves,\inst{2}  
\and
Breno R. L. Galvão\inst{2,5}
          }

   \institute{
             Dept. Ciencias Integradas, Facultad de Ciencias Experimentales, Centro de Estudios Avanzados en Física, Matemática y Computación,Unidad Asociada GIFMAN, CSIC-UHU, Universidad de Huelva, Spain \email{edgar.mendoza@dci.uhu.es}
    \and
             Centro Federal de Educação Tecnológica de Minas Gerais, CEFET-MG. 
Av. Amazonas 5253, 30421-169, Belo Horizonte, Minas Gerais, Brazil
              \email{brenogalvao@gmail.com}
\and
Instituto Universitario "Carlos I" de Física Teórica y Computacional, Universidad de Granada, Spain
\and
Instituto de Pesquisa e Desenvolvimento (IP\&D), Universidade do Vale do Para{\'i}ba (UNIVAP), Av. Shishima Hifumi 2911, São
José dos Campos, 12244-390, Brazil
\and
Department of Chemistry and Chemical Biology, University of New Mexico,
Albuquerque, 87131, New Mexico, USA.
             }

\titlerunning{New SiS destruction and formation routes via neutral-neutral reactions.}
\authorrunning{Mendoza {\em et al.}}

   \date{Received --; accepted --}

 
  \abstract
   {Among the silicon bearing species discovered in the interstellar medium, SiS and SiO stand out as key tracers due to their distinct chemistry and variable abundances in interstellar and circumstellar environments. Nevertheless, while the origins of SiO are well documented, the SiS chemistry remains relatively unexplored.}
   {Our objective is to enhance the network of Si- and S-bearing chemical reactions for a gas-grain model in molecular clouds, encompassing both low and high metallicities. To achieve this, we have calculated the energies and rate coefficients for six neutral atom-diatom reactions involved in the SiCS triatomic system, with a special focus on the C+SiS and S+SiC collisions.}
   {We employ the coupled cluster method with single and double substitutions and a perturbative treatment of triple substitutions  (CCSD(T)) refined at the explicitly correlated CCSD(T)-F12 level. With these computational results in conjunction with supplementary data from the literature, we construct an extended network of neutral-neutral chemical reactions involving Si- and S-bearing molecules. To assess the impact of these chemical reactions, we performed time-dependent models employing the Nautilus gas-grain code, setting the gas temperature to 10~K and the H$_2$ density to 2$\times$10$^4$ cm$^{-3}$. The models considered two initial abundance scenarios, corresponding to low and high metallicity levels. Abundances were computed using both the default chemical network and the constrained network, enriched with newly calculated reactions. }
   {The temperature dependence for the reactions involving SiS were modelled to the $k(T)=\alpha \left( T/300 \right)^{\beta} \exp{(-\gamma/T)}$ expression, and the coefficients are provided for the first time. The high-metallicity models significantly boost the SiS production, resulting in abundances nearly four orders of magnitude higher compared to low-metallicity models. Higher initial abundances of C, S, and Si, roughly~$\sim$~2, 190, and 210 times higher, respectively, contribute to this. Around the age of 10$^3$ yr, destruction mechanisms become relevant, impacting the abundance of SiS. The proposed production reaction S + SiC $\longrightarrow$ C + SiS, mitigates these effects in later stages. By expanding the gas reaction network using a high metallicity model, we derived estimates for the abundances of observed interstellar molecules, including SiO, SO and SO$_2$.}  
   {We demonstrate the significance of both SiC+S and C+SiS channels in the SiS chemistry. Notably, the inclusion of neutral-neutral mechanisms, particularly via Si+HS and S+SiC channels, played a pivotal role in determining SiS abundance. These mechanisms carry a significance on a par with the well-known and fast ion-neutral reactions.}
 
\keywords{Astrochemistry --
        Molecular data -- Molecular processes -- 
        ISM: abundances
               }

   \maketitle
%

\section{Introduction}

The study of S (sulphur) and Si (silicon) atoms and related molecules in the interstellar medium (ISM) is of great importance in astrophysics, astrochemistry and also astrobiology. For example, these species are used as:
i) Tracers of cosmic dust: S and Si are present in interstellar dust, an important component of the ISM. By examining the abundance and distribution of these atoms, we can gain insights into the properties of interstellar dust, which is believed to play a crucial role in star formation~\citep{DOD21:7003}. ii) Indicators of interstellar chemistry: S and Si are also important tracers of the chemical processes occurring in the ISM. For instance, the presence of specific molecular ions, such as \ce{SiH+}, can provide clues about the chemical reactions via radiative association of Si with H$_3^+$ in the surrounding gas of protostellar objects or in the diffuse interstellar medium in general  ~\citep{DEA78:415,LEA85:353}. iii) Probes of interstellar physical conditions: The abundances of sulfur and silicon ions, relative to molecular hydrogen, offer valuable insights into the chemical reactions responsible for their formation, as well as the temperatures involved in these processes \cite{Stancil2000}. iv) Tracers of the stellar evolution: The presence of these elements in protostellar environments can shed light on the processes leading to the formation of planetary systems. Silicates are thought to be components of the rocky planets in our own solar system~\citep{DAV20:e2019JE006227}. v) Paramount in the studies of astrobiology. Molecules containing sulphur play a crucial role in the synthesis of fundamental amino acids, such as methionine, which is an important building block for proteins~\citep{Parker2011}.  Overall, the study of S and Si atoms, along with the resultant molecules from them, provides relevant clues on the evolution and molecular complexity across the different evolutionary stages of the ISM \citep{Herpin2009}.

Sulfur and silicon are among the most abundant elements in the universe. In particular, the study of silicon monosulfide (SiS) is especially significant in the context of neutral-neutral reactions under interstellar conditions. Surveys have been dedicated to the search of interstellar Si- and S-bearing molecules; however, detecting SiS is more challenging due to its expected low abundance. Nonetheless, these compounds provide valuable insights into the temperature, density, and ionisation state of the gas in regions such as circumstellar shells. Additionally, they serve as a good indicator of the molecular complexity in space environments \citep{Riviere2019,Massalkhi2019study,Massalkhi2020}.

In this study, we focus on neutral-neutral reactions involving the SiS molecule. On the one hand, in the context of the chemical building blocks of life,  this is particularly relevant for the formation of organic molecules containing sulfur, e.g., amino acids as cysteine and methionine \citep{Todd2022}. On the other hand, 
SiS is also thought to play a role in the formation of dust in carbon-rich star envelopes and may contribute to grain growth through molecular condensation and accretion mechanisms~\citep{Massalkhi2019study}. 
Its reactions with other molecules, such as \ce{C2H2} and \ce{NH3}, are believed to be one of the main pathways for the formation of silicon-based dust grains~\citep{ACC21:44}.

The first detection of SiS was reported by \citet{morris1975detection} through the J=5--4 and 6--5 transitions at frequencies $\sim$~90.771 and 108.924~GHz, respectively, toward the evolved carbon-rich assymptotic giant branch (AGB) star IRC+10216.  
Subsequent studies have successfully reported the presence of SiS in the circumstellar envelopes of both evolved carbon- and oxygen-rich AGB stars~\citep{exposito2006high,Schoier2007}. Therefore, SiS formation likely occurs within conditions akin to circumstellar envelopes, outflows, bow shocks, and the expanding bubbles surrounding stellar objects. These environments, characterized by thermal and radiative processes, are favorable to the triggering of the chemical reactions responsible for the emergence of SiS  ~\citep{Ziurys1988,agundez2012molecular,fonfria2015abundance,PRI15:L13,Ortiz2023}. SiS has also been detected in various other environments, including
star-forming regions as Sgr B2 and Orion~KL, as well as in the protostellar L1157-B1~\citep{DIC81:112,Ziurys1988,ZIU91:260,TER11:A26,POD17:L16}. 
In external galaxies, the detection of SiS has not yet been confirmed, although it is expected in the near future. Approximately 30\% of the molecules detected in our Galaxy have also been identified in external galaxies \citep{McGuire2022}. Regarding Si-bearing molecules, the first detection in the dense gas of nearby galaxies was reported for the molecule of SiO by \citet{Mauersberger1991}.

The abundances of SiS vary among interstellar sources, including oxygen-rich and carbon-rich evolved stars. To address this issue in the context of astrochemical models, it is crucial to have values for the elemental abundances of the source~\citep{vanDishoeck2013}, specifically, the metallicity, which represents the abundance of elements heavier than hydrogen and helium~\citep{Kobayashi2011}. It is also essential to understand key chemical reactions, both in the gas-phase and assisted by grains, to establish a solid correlation between predicted and observed abundances.
It is worth noting that whereas astrochemical models often focus on the high metal abundances, which are more aligned with solar and Galactic values, low-metal models have received less attention. However, the latter are crucial to understand the chemistry in environments with low-metal content \citep{Shimonishi2021}. The increasing detection of molecules in the outer Galaxy, the Galactic Halo and nearby galaxies highlights the necessity for further studies considering low-metallicity regions \citep{Salvadori2010,Acharyya2018,Fontani2022}.
Thus, this work analyses both low and high metallicity environments and their influence on the molecular abundances of SiS and other Si-bearing species. 
Recent research on interstellar molecules in nearby low-metallicity galaxies indicates that metallicity profoundly influences the chemistry of star-forming cores. In a study by \citet{Shimonishi2021}, a hot molecular core was identified in the extreme outer galaxy, which was associated with high-velocity SiO outflows. Their findings highlight the presence of significant molecular complexity even in the primordial conditions of the extreme outer Galaxy.

\cite{Wakelam2010} investigated rigorously the impact of various physical and chemical parameters on astrochemical models of dense clouds. Their findings highlighted that reaction rate coefficients and initial chemical abundances were the most critical parameters, while gas temperatures and densities had a relatively lesser impact. \citet{Gerner2014} performed a study about the chemical evolution in the early stages of massive star formation. As they explored the realm of the initial chemical abundances, a set of low metal values was used in order to effectively mimic the chemistry of infrared dark clouds. \citet{Bialy2015} carried out a study on the chemistry of dense interstellar gas, examining the impact of varying metallicities and ionisation parameters. Their study highlighted the significance of ion-molecule processes within Galactic conditions characterised by exceptionally low metal abundances. Studies of photodissociation regions through chemical models have revealed that some organic molecules are not accurately modelled by astrochemical codes. This issue has been addressed by testing variations in the C/O ratios of the elemental gas-phase abundances  (\citealt{Romane2019} and references therein). Regarding the metallicity of the Milky Way, it is typically high in the galactic center and decreases with distance from it~\citep{Wilson1992,Chiappini2001}. High values are expected in the spiral arms, while lower levels are expected in the halo. Some studies have focused on depicting the integrated metallicity profile of our galaxy~\citep{Lian2023}. In nearby galaxies, star-forming regions serve as excellent targets to study the interstellar chemistry in low-metallicity extragalactic regimes, such as the case of the Magellanic clouds~\citep{Esteban2022}.

The lack of experimental reactions data for SiS in the gas phase makes the computational predictions a powerful tool for bridging the gap between laboratory studies and observations. These predictions also contribute to enhancing ISM reaction databases with reactions energies and rate coefficients.
In previous works, calculations have demonstrated that SiS can be rapidly destroyed in collisions with atomic oxygen~\citep{ZAN18:38,PAI18:1858} and can also be formed from collisions between atomic silicon and simple sulphur bearing species~\citep{PAI20:299,DOD21:7003}. Subsequently, gas-grain astrochemical models  applied to L1157-B1 has shown that such neutral-neutral reactions can change the predicted abundances of SiS by more than one order of magnitude~\citep{MOT21:37,CAM22:369}.

In the literature, a recent summary of the results regarding the reactions between neutral Si- and S-bearing molecules has been reported in Table~1 of the work by \cite{CAM22:369}. These results are important for the astrochemical modelling of such species. In that work, it was shown that SiS is stable with respect to collisions with H, \ce{H2}, and \ce{O2}, but it is quickly depleted by atomic oxygen. However, atomic carbon is also an abundant species which could contribute to the SiS depletion. This channel has not been investigated yet and may cause a large impact on the astrochemical models. Considering this, the primary objective of this study is to provide computational insights into these reactions and their reverses:
\begin{align*}
 \rm C(^3P) + SiS(^1\Sigma) \rightarrow & \rm S(^3P) + SiC(^3\Pi)\\
 \rightarrow & \rm  Si(^3P)  + CS(^1\Sigma)
 \end{align*}

Furthermore, our investigation extends to addressing questions about gas phase chemical processes. With this aim, we simulate molecular cloud conditions and compare scenarios involving both high and low elemental abundances. This approach is particularly significant to enhance our understanding of chemistry in environments characterised by a poor metal content, such as the galactic halo and external galaxies.

The structure of this article is organised as follows: In Section~\ref{sec:method} we describe the methodology employed for electronic structure calculations related to the SiS+C channel and the gas phase astrochemical model. Section~\ref{sec:results} presents the computational results, delving into detailed discussions regarding the reaction products, mechanisms, and reaction rates. In Section~\ref {sec:modelling} we present and discuss the results concerning the gas phase chemical models for environments with both low and high metallicities. Moreover, we provide a compilation of updated reactions associated with SiS, thereby expanding and enhancing the chemical model. Finally, in Section~\ref{sec:conc}, we provide our concluding remarks derived from this study.

\section{Methods}
\label{sec:method}

\subsection{Electronic Structure calculations}

To investigate the possible outcomes of the \ce{C(^3P) + SiS(^1\Sigma)} reactions we have explored the triplet state potential energy surface of the CSiS triatomic molecule and its dissociation channels. Note that the singlet electronic state is not correlated with the ground state reactants. For a preliminary exploration of the several minima and transition states, we have first employed unrestricted DFT calculations with the M06-2X functional \citep{ZHA08:215} with the aug-cc-pV($T$+$d$)Z basis set~\citep{DUN01:9244}.

For accurately obtaining the energies involved in the reactions, we re-optimised all structures using the coupled cluster method with single and double substitutions and a perturbative treatment of triple substitutions  (CCSD(T))~\citep{KNO93:5219} which was followed by vibrational frequencies 
calculations at the same level, and a further single point energy refinement at the explicitly correlated CCSD(T)-F12 method~\citep{ADL07:221106,KNI09:054104}, with the cc-pVQZ-F12 basis set~\citep{PET08:084102}. All reported results are zero-point energy (ZPE) corrected and denoted using the standard quantum chemistry notation as CCSD(T)-F12/cc-pVQZ-F12//CCSD(T)/cc-pVTZ+ZPE(CCSD(T)/cc-pVTZ).
All calculations were performed with the MOLPRO 2015 package~\citep{MOLPRO_brief}.

\subsection{Chemical modelling}

Astrochemical gas-grain models were carried out with the aim of understanding how the incorporation of new and previously investigated reactions affects the relative abundances of elements C and S, along with compounds such as SiS and SiC. To achieve this, we employed the {\sc Nautilus} gas-grain code \citep{Ruaud2016} to simulate the chemical processes that occur in interstellar environments. Through Nautilus, we computed time-dependent models aimed at determining the relative abundances of the aforementioned species. 

{\sc Nautilus} code uses the so-called kida.uva.2014 database, a comprehensive compilation of chemical reactions and their corresponding rate coefficients that are relevant for simulating chemical processes taking place in astronomical environments, including molecular clouds and in general, the interstellar medium. The database includes around 489 species, formed by 13 elements, and incorporates 7509 reactions \citep{KIDA}. Among these chemical reactions, there are around 10 chemical reactions involving SiS, with detailed entries available in the KIDA database.\footnote{Kinetic Database for Astrochemistry (KIDA) \url{https://kida.astrochem-tools.org/}} In this study, our methodology focuses on contrasting the SiS abundances calculated using the default KIDA chemical network with those resulting from the integration of the new chemistry presented herein. Two specific objectives were pursued to examine the SiS chemistry. Firstly, six newly computed reactions were introduced along with their corresponding rate coefficients. Secondly, 11 chemical reactions were revised from the literature and updated to compute an extended model. The outcomes were carefully compared with the results obtained from the default chemical network.

To describe the physical conditions under which these chemical reactions are computed, we adopted the conventional conditions of molecular clouds as a modelling template \citep[e.g.,][]{vidal2017reservoir}.  These cold interstellar objects exhibit gas temperatures between $\sim$10 K and 30~K, with densities ranging from $\sim$~10$^2$ to 10$^4$ cm$^{-3}$. They are predominantly composed of H$_2$ and CO, among other molecules relevant to our study. Visual extinction can vary, often reaching $A_V \thickapprox$ 10--50 mag due to density and opacity factors \citep{Lada2003}. The cosmic ray ionisation rate ranges from a few of 10$^{-17}$~s$^{-1}$ to a few of 10$^{-16}$~s$^{-1}$ \citep{Padovani2009}, although for our work, we adopted the standard value $\zeta =$1.3 $\times$ 10$^{-17}$ s$^{-1}$.

In terms of the chemical properties, our simulations are based on time-dependent gas models that initiate with the elemental abundances outlined in Table~\ref{table1}. Thus, our primary focus is to study the gas formation and destruction routes of SiS and SiC. Regarding the initial abundances, we investigated the chemistry of SiS and SiC considering two distinct regimes that characterize the regions exhibiting varying levels of metallicity \citep{Acharyya2015}, encompassing both low and high values. As was highlighted in the Introduction, low-metallicity regions are commonly situated within the outer halo of the Milky Way, while high-metallicity regions tend to be concentrated in the central bulge or along the spiral arms of the galactic disc. In the Table~\ref{table1} we designate LM and HM as low- and high-metallicity models, respectively. This convention will be used from this point  forward. Thus, considering  the initial elemental composition as a starting point and let it evolve over time using a scale of  1$\times$10$^7$~yr, the abundances of molecules as SiS are determined via the intricate interplay within the network of chemical reactions. The abundance estimates allow us to evaluate the balance between production and destruction reactions over the course of time. 

\begin{table}
	\caption{Initial abundances computed in the chemical model for low- and high-metal abundance scenarios (LM and HM) in molecular clouds. Values have been collected from \citet{Acharyya2015} and references therein.}
	\label{table1}
    \centering
	\begin{tabular}{lll}
		\hline
		\textbf{Species} & \textbf{LM} & \textbf{HM}\\
		\hline
		He     & 9.0  $\times$ 10$^{-2}$    &  9.0  $\times$ 10$^{-2}$ \\
		C      & 7.3 $\times$ 10$^{-5}$     &  1.2 $\times$ 10$^{-4}$ \\
		O      & 1.76 $\times$ 10$^{-4}$     & 2.56 $\times$ 10$^{-4}$ \\
        N      & 2.14 $\times$ 10$^{-5}$     & 7.60 $\times$ 10$^{-5}$ \\
        Cl      & 1.0 $\times$ 10$^{-9}$     &  1.80 $\times$ 10$^{-7}$ \\
        Fe      & 3.0 $\times$ 10$^{-9}$     &  2.0 $\times$ 10$^{-7}$ \\
        Mg      & 7.0 $\times$ 10$^{-9}$     &  2.40 $\times$ 10$^{-6}$ \\
        Na      & 2.00 $\times$ 10$^{-9}$     & 2.00 $\times$ 10$^{-7}$ \\
        S      & 8.00 $\times$ 10$^{-8}$     &  1.50 $\times$ 10$^{-5}$ \\
        Si      & 8.00 $\times$ 10$^{-9}$     & 1.70 $\times$ 10$^{-6}$ \\
  \hline
	\end{tabular}
\end{table}

\section{Computational  Results}
\label{sec:results}

\subsection{Reaction products}

High quality electronic structure calculations at the level of theory  CCSD(T)-F12/cc-pVQZ-F12//CCSD(T)/cc-pVTZ+ZPE(CCSD(T)/cc-pVTZ) were carried out for the possible reaction products. These calculations allowed us to estimate the reaction energies

\begin{align}
\label{eq:Rp1}  \rm C(^3P) + SiS(^1\Sigma) \rightarrow & \rm S(^3P) + SiC(^3\Pi) \,\,\,\,\,\, \rm +191 \,kJ\,mol^{-1}\\
\label{eq:p2}  \rightarrow & \rm  Si(^3P)  + CS(^1\Sigma)\,\,\,\,\,\,\,\, \rm -98 \,kJ\,mol^{-1} 
\end{align}

From the C+SiS reactants, only the Si+CS products  (Eq.~\ref{eq:p2}) can be obtained exothermically. This is due to a deeper potential energy well in the CS diatomic molecule if compared to SiS. The formation of SiC is significantly endothermic  (Eq.~\ref{eq:Rp1}) and, therefore, it should not be a possible route in the low temperature regions of the interstellar medium (ISM). 

We also predicted that silicon carbide (SiC), which has also been observed in the ISM~\citep{ZIU06:12274}, may be destructed by neutral S atoms through exothermic processes yielding C+SiS (the reverse of  Eq.~\ref{eq:Rp1}, or Eq.-~\ref{eq:Rp1}) with an energy release of $\rm -191 \,kJ\,mol^{-1}$, or through the reaction

\begin{align}
\label{eq:P2p1} \rm S(^3P) + SiC(^3\Pi) \rightarrow & \rm Si(^3P)  + CS(^1\Sigma) \rm  \,\,\,\,\,\, \rm -289 \,kJ\,mol^{-1}
\end{align}

Finally, carbon monosulfide (CS) is thermodynamically stable with respect to collisions with atomic silicon. 

\subsection{Reaction mechanisms}

  \begin{figure*}
   \centering
   \includegraphics[width=0.75\textwidth]{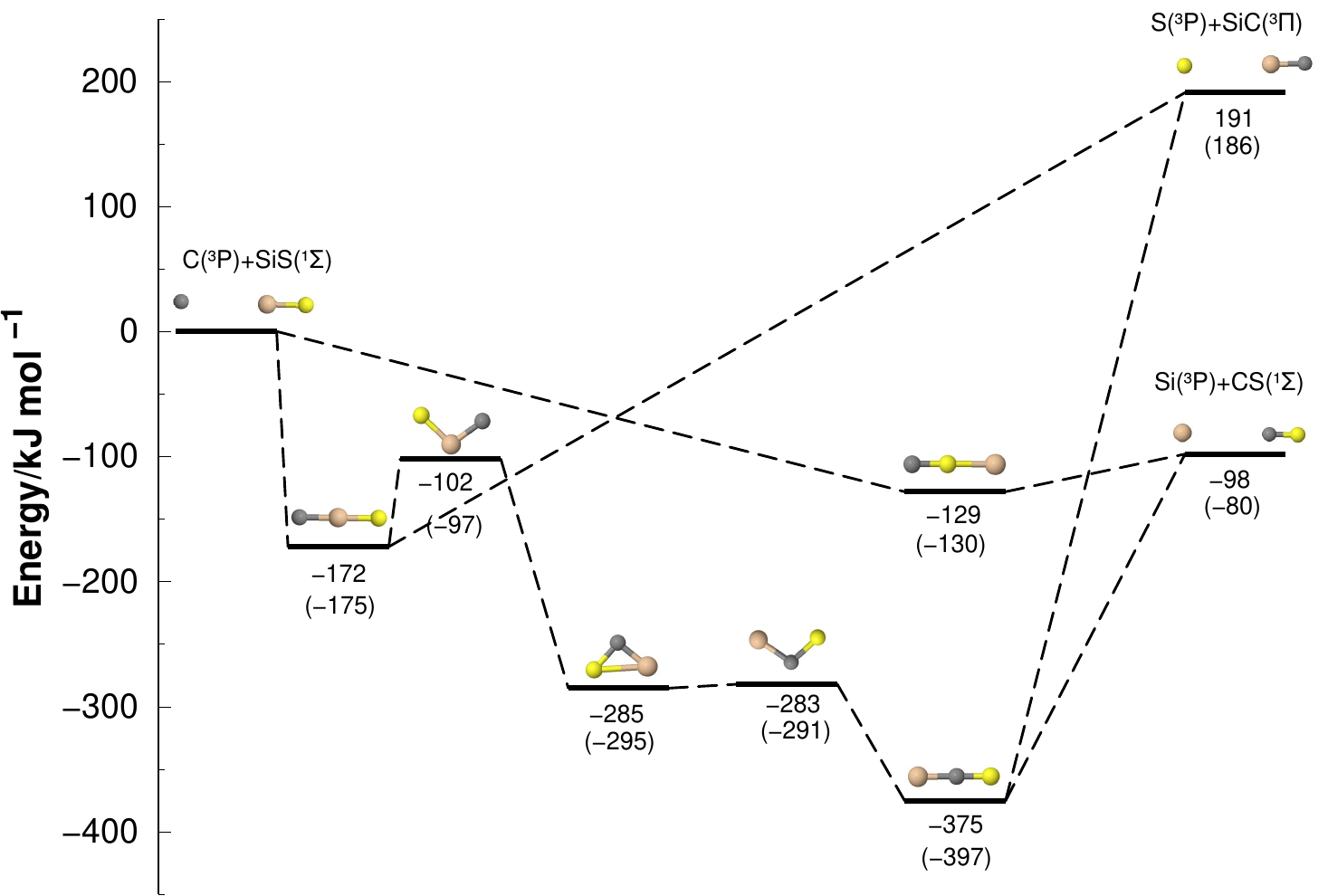}
      \caption{Potential energy diagram for the $^3$CSiS system and its dissociation channels. Plain numbers refer to the CCSD(T)-F12/cc-pVQZ-F12//CCSD(T)/cc-pVTZ+ZPE(CCSD(T)/cc-pVTZ) results, whereas those in parenthesis were obtained with the exploratory M06-2X/cc-pVTZ calculations. All results are ZPE corrected and given relative to C+SiS.}
         \label{Fig:PES}
   \end{figure*}

Even though we have identified exothermic reactions in the previous section, their potential energy barriers could be significant, which would turn them kinetically unfeasible in the cold parts of the ISM. In fact, most reactions included in the astrochemical databases in the past were ion-molecule ones, as neutral-neutral reactions often present reaction barriers. However, it has been shown that  not only several  neutral-neutral reactions may be kinetically viable, but they also have a large impact on the abundances of several molecules~\citep{DOD21:7003, MOT21:37, CAM22:369, GOM23:e011}. For this reason, apart from thermochemical considerations, one must also look into the potential energy surface for the reactions and assess the kinetic viability. The results for the present chemical system are summarised in Fig.~\ref{Fig:PES}.

It can be noted in Fig.~\ref{Fig:PES} that when the C atom approaches SiS, barrierless abstraction of the sulphur atom may happen
to yield CS passing through a shallow CSSi minimum. Additionally, the atomic carbon can approach the diatomic SiS via the Si atom to form CSiS, which will, in turn, either isomerise or  come back to the reactants. Ultimately, these two channels will lead to Si+CS, although the latter  mechanism will imply a longer lifetime and energy randomisation among the degrees of freedom of the triatomic complex. Most importantly, no energy barrier is observed in any entrance or exit channel. This implies that the intermediate step of atom-diatom capture  will dominate the reactivity and, thus, determine the reaction rate coefficient values.

Figure~\ref{Fig:PES} also shows the existence of different configurations of the triatomic molecule in its triplet state: three linear minima, two bent and a shallow cyclic structures. The most strongly bonded species is SiCS, which exhibits the global minimum of the potential energy surface (PES). The  minimum of the cyclic configuration is very shallow, and the energy required to produce the isomer SiCS is very low, lying below the accuracy of the quantum chemical calculations. Nevertheless, the existence of such intermediate structure is unlikely to influence the rate coefficient values.

In the S+SiC reaction, the sulphur atom can easily abstract a Si atom to yield SiS+C, or  the carbon atom to product Si+CS. Both pathways are barrierless and do not  undergo any isomerisation reaction. Therefore, we can anticipate that the rate coefficient in the  S+SiC collision will be branched roughly equally between the products Si+CS and C+SiS.

\subsection{Reaction rates}

A rigorous theoretical approach for calculating the rate coefficients for all reactions discussed here would be to fit thousands of {\em ab initio} energies to obtain an analytic representation of the PES, which could be followed by quasi-classical trajectories simulations to obtain reliable results. This is, however, a very time consuming process. Nevertheless, it is generally possible to guess that exoergic and barrierless reactions  (such as Eq.~\ref{eq:p2}, Eq.-\ref{eq:Rp1} and Eq.~\ref{eq:P2p1}) have rate coefficients of a few $10^{-10} \rm cm^3 s^{-1}$ not significantly depending on the temperature. Such reactions are the most important neutral-neutral ones involving Si to drive the chemistry in the cold environments of the ISM.  To obtain a first estimate on the reactions studied in this work, we employ simple statistical methods to calculate the rate coefficients required for the chemical modelling. We have employed the MESS software package~\citep{GEO13:12146}  to calculate rigid rotor/ harmonic oscillator rate coefficients using standard transition state theory (TST), including the Eckart tunnelling correction~\citep{ECK30:1303}. For the barrierless capture steps, we have followed an approach similar to that of ~\cite{CON21:169}, in which we fit the {\em ab initio} energies over the minimum energy path as a function of the internuclear separation ($r$) to a simple  $-C_6/r^6$ function to obtain $C_6$. This value is used in a phase space theory (PST) calculation~\citep{PEC65:3281,CHE86:2615} as implemented in MESS.

\begin{figure}
   \centering
 \includegraphics[width=0.9\linewidth]{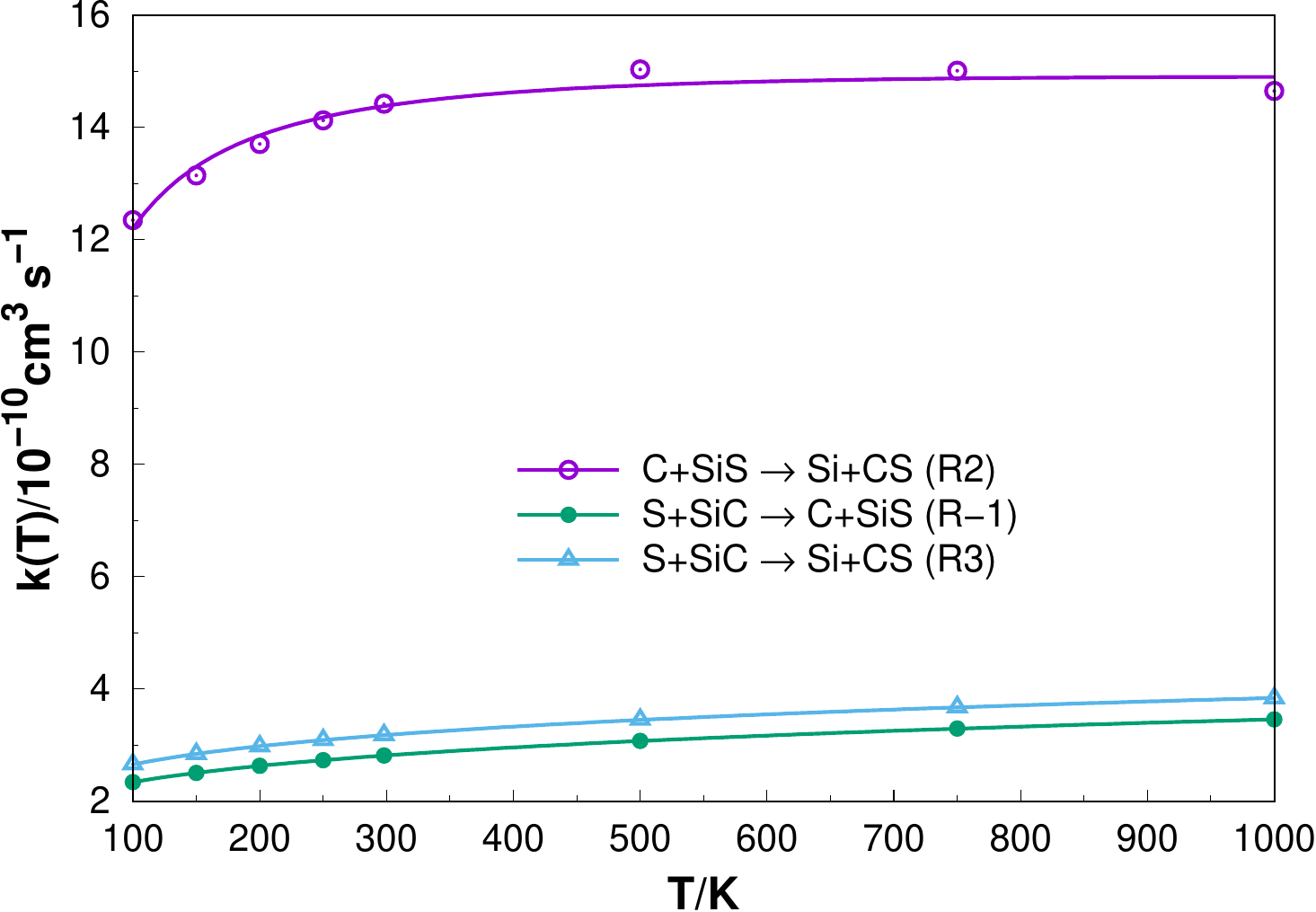}
      \caption{Calculated rate coefficients for the  exoergic and barrierless reactions.       }
         \label{Fig:fast}
   \end{figure}

The results for the exoergic and barrierless reactions are given in Fig.~\ref{Fig:fast}. As expected, the barrierless reactions do not show a significant temperature dependence. Interestingly, the rate coefficients for the destruction of SiS by atomic carbon  (Eq.~\ref{eq:p2}) are substantially higher than the others. This can be rationalised by two aspects: (i) in the C+SiS collision, only one exoergic product is accessible, while the S+SiC one is branched into two possibilities (see Fig.~\ref{Fig:PES}) and (ii) the large difference in reduced mass between different atom-diatom pairs~\citep{LIG64:3221}. 

The two possible outcomes of the S+SiC collisions  (Eq.-\ref{eq:Rp1} and Eq.~\ref{eq:P2p1}) are branched into similar values of the rate coefficients   through direct abstraction mechanisms. The products are determined by the angle of attack of the incoming S atom: the diatomic species SiS is formed if sulphur approaches the silicon atom, while if it approaches C,  the species CS is produced.

\begin{figure}
   \centering
 \includegraphics[width=0.9\linewidth]{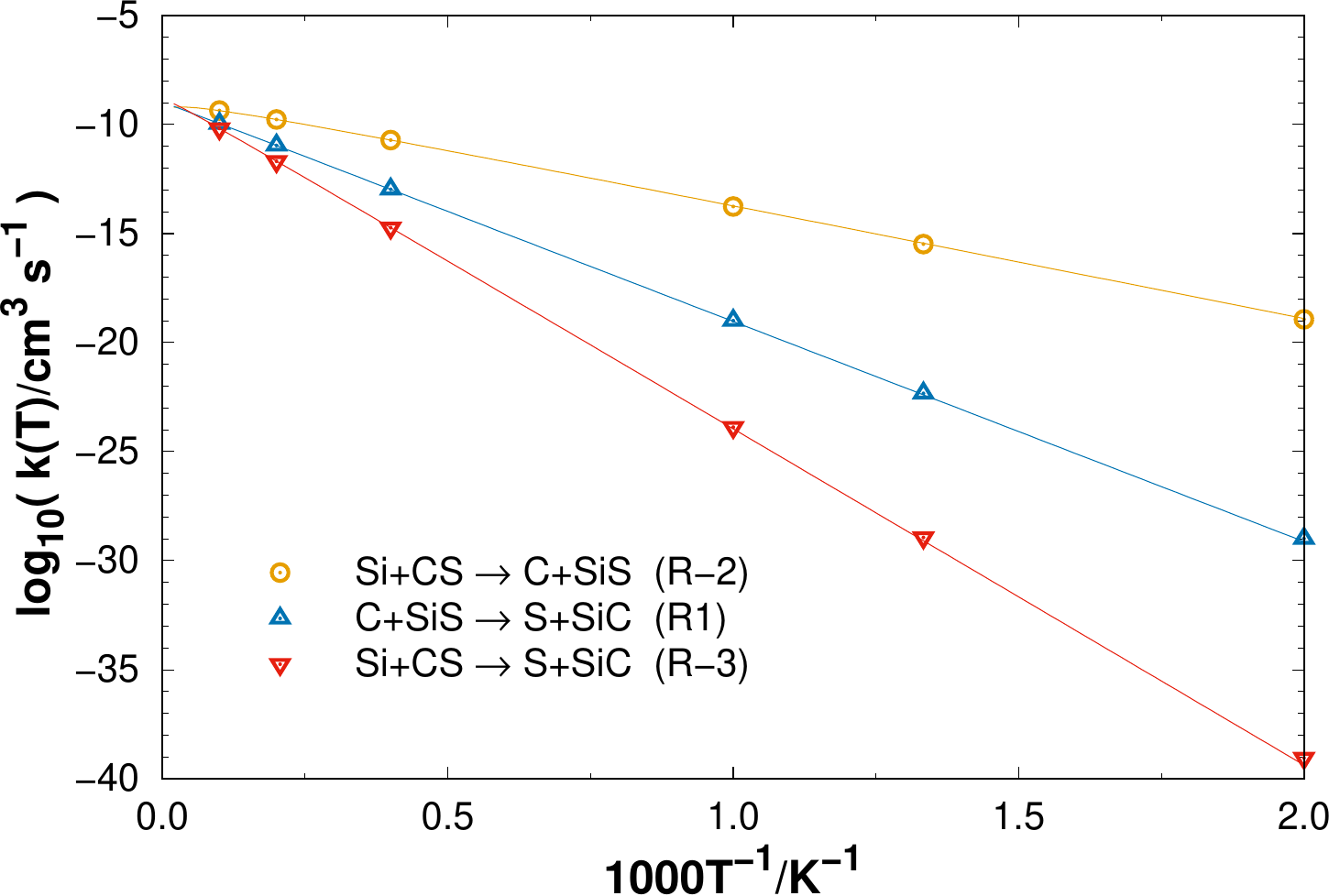}
      \caption{Calculated rate coefficients for the endoergic reactions. }
         \label{Fig:slow}
   \end{figure}

The rate coefficients for the endoergic reactions are given in Fig.~\ref{Fig:slow} in an Arrhenius plot, where the slope is related to the activation energy. At 1000~K, the highest rate coefficient (Eq.-\ref{eq:p2}) is only $1.7\times 10^{-14} \rm cm^3 s^{-1}$ and only at 5000\,K does it reach the order of magnitude of $10^{-10} \rm cm^3 s^{-1}$. Therefore, such reactions are unlikely to play a role in the evolution of silicon and sulphur-bearing molecules in cold regions. 

The values obtained for the rate coefficients for all 6 reactions have been fitted to the modified Arrhenius form:

\begin{equation*}
\label{eq:modified-arrhenius}  
k(T)=\alpha \left( T/300 \right)^{\beta} \exp{(-\gamma/T),}
\end{equation*}

the fit parameters corresponding to the $\alpha$, $\beta$  and $\gamma$ constants are  gathered in Table~\ref{tab:net}. For completeness, we also include in this table the results from previous works, to provide a comprehensive set of reaction rates to be used in the modelling of the abundances of Si- and S-bearing molecules.

\begin{table*}
	\caption{Rate coefficients expressed as $k(T)=\alpha \left( T/300 \right)^{\beta} \exp{(-\gamma/T)}$ for the neutral-neutral reactions involving silicon and sulphur.}
	\label{tab:net}
    \centering
	\begin{tabular}{rlcccc}
		\hline
	index$^a$	& Reaction & $\alpha \rm (cm^{-3}s^{-1})$ & $\beta$ & $\gamma \rm (K)$ & description$^b$\\
		\hline

1 & \ce{C +SiS}$\rightarrow$\ce{S +SiC}  & $1.20\times 10^{-9}$ & 0.020 & 23261 &This work  \\
-1 & \ce{S +SiC}$\rightarrow$\ce{C +SiS} & $2.81\times 10^{-10}$ & -0.171 & -0.383 & This work\\
\hline
2 & \ce{C +SiS}$\rightarrow$\ce{Si +CS} & $1.59\times 10^{-9}$ & 0.0277 & 29.25 & This work  \\
-2 & \ce{Si +CS}$\rightarrow$\ce{C +SiS} & $5.74\times 10^{-9}$ & 0.389 & 12174 & This work\\
\hline
3 & \ce{S +SiC}$\rightarrow$\ce{Si +CS} & $3.20\times 10^{-10}$ & -0.152 & 2.073 & This work\\
-3 & \ce{Si +CS}$\rightarrow$\ce{S +SiC} & $3.84\times 10^{-9}$ & 0.143 & 35604 & This work\\*[1mm]
\hline

4& \ce{Si + HS}$\rightarrow$ \ce{SiS + H} & $0.916\times 10^{-10}$ & -0.6433 & 11.525 & QCT,~\citep{MOT22:3555} \\
-4& \ce{SiS +H}$\rightarrow$ \ce{Si + HS} & 0 & 0 & 0 & Endo.,~\citep{ROS18:695} \\
\hline

5& \ce{Si +H2S}$\rightarrow$ \ce{SiS + H2} & $1\times 10^{-10}$ & 0 & 0 & Exp.,~\citep{DOD21:7003} \\		
-5& \ce{SiS +H2}$\rightarrow$ \ce{Si + H2S} & 0 & 0 & 0 & Endo.,~\citep{PAI20:299} \\	
\hline

6 & \ce{Si +SO}$\rightarrow$ \ce{SiS + O} & $1.77\times 10^{-11}$ & 0.16 & -20 & QCT,~\citep{ZAN18:38} \\

7 & \ce{Si +SO}$\rightarrow$ \ce{SiO + S} & $1.53\times 10^{-10}$ & -0.11 & 32 & QCT~\citep{ZAN18:38} \\	

8 & \ce{SiS +O}$\rightarrow$ \ce{SiO + S} & $9.53\times 10^{-11}$ & 0.29 & -32 & QCT~\citep{ZAN18:38} \\

9 & \ce{SiH +S}$\rightarrow$ \ce{SiS + H} & $ 0.63\times 10^{-10}$ &  -0.11 & 11.6  & QCT,~\citep{GAL23:000} \\	
9b & \ce{SiH +S}$\rightarrow$ \ce{SH + Si} & 0.025$\times 10^{-10}$ & -0.13  & 9.38 & QCT,~\citep{GAL23:000} \\	

10 & \ce{SiH +S2}$\rightarrow$ \ce{SiS + SH} & $1\times 10^{-10}$ & 0 & 0 & PES,~\citep{ROS18:695} \\

11 & \ce{Si +SO2}$\rightarrow$ \ce{SiO + SO} & $1\times 10^{-10}$ & 0 & 0 & PES,~\citep{CAM22:369} \\	

12 & \ce{SiS +O2} & 0 & 0 & 0 & Endo.,~\citep{CAM22:369} \\	
		\hline
	\end{tabular}
	
\tablefoot{$^a$~Negative numbers indicate the inverse reactions.
$^b$~References and description of the approaches used to obtain the results:  QCT stands for\ ``computed by quasiclassical trajectories simulations''; Exp. stands for  ``inferred from experiments''; PES stands for ``inferred from the absence of a potential energy barrier''; Endo stands for ``considered not to occur due to endoergic character''.}
\end{table*}

\section{Astrochemical model results}
\label{sec:modelling}

\subsection{Models including the new reactions}\label{sec4.1}

We conducted several model computations  to study the chemical evolution under the conditions representative of interstellar molecular clouds. In particular, we pay  special attention to the impact of altering the initial chemical abundances during the time-dependent simulation.
In order to reproduce the physical conditions of a molecular cloud, we employed standard values for gas density and temperature, which were set at  $T$=10~K and $n_{\text H}$=2 $\times 10^4$ cm$^{-3}$, respectively. Furthermore, we adopted a visual extinction of $A_V$=10~mag and a cosmic rays ionisation rate of $\zeta$=1.3$\times$10$^{-17}$ s$^{-1}$. Taking into account that the code includes surface mechanisms, we assumed an equal temperature of 10~K for both the gas and dust components. For characterising the dust properties, we utilised a single grain radius of 0.1 $\mu$m, a dust-to-gas mass ratio of 0.01, and a dust bulk density of 3~g cm$^{-3}$~\citep{Hasegawa1993}. As in this work we primarily focus on the study of the variation of the chemical properties, the physical parameters were set to the standard conditions of a molecular cloud. Thus, our main adjustments were made to the chemical aspects, specifically the initial chemical abundances (Table~\ref{table1}) and the chemical reactions.
Regarding the chemical reaction network, we employed the kida.uva.2014 catalogue of chemical reactions~\citep{KIDA} and supplemented it with the computed reactions analysed in this study. The chemical equations and kinetic coefficients associated with these reactions are provided in Table~\ref{tab:net}. This approach plays a pivotal role in evaluating the impact of such reactions within the framework of models that include high and low metallicity abundances.

Considering the specified physical conditions, we computed our models using two different sets of initial chemical abundances. These sets were intended to simulate both low- and high-metallicity environments, and their values are shown in Table~\ref{table1}. Comparing the abundances in Table~\ref{table1} between the LM and HM scenarios, it can be observed that, from a HM environment to a LM one, the elements abundances undergo a  trend downward  with factors of approximately 2, 190 and 210 observed for the elements C, S and Si, respectively. This significant depletion directly affects the formation and destruction pathways of the SiS molecule and related species studied here.

\begin{figure*}
\begin{center}
\includegraphics[width=\columnwidth,keepaspectratio]{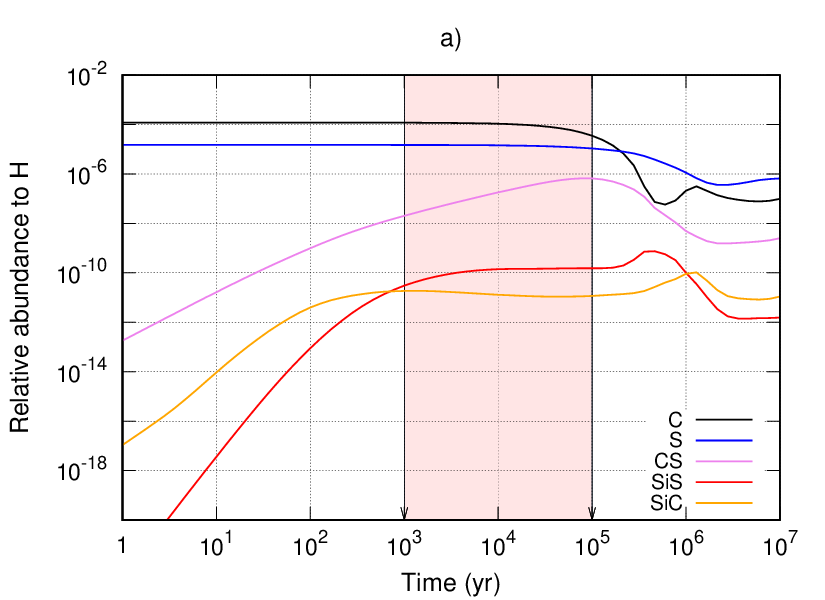}
\includegraphics[width=\columnwidth,keepaspectratio]{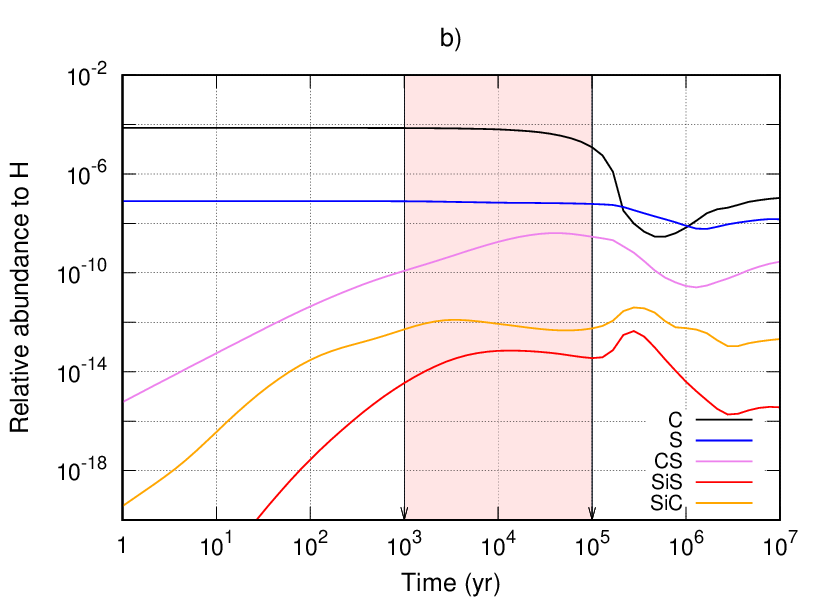}\\
\includegraphics[width=\columnwidth,keepaspectratio]{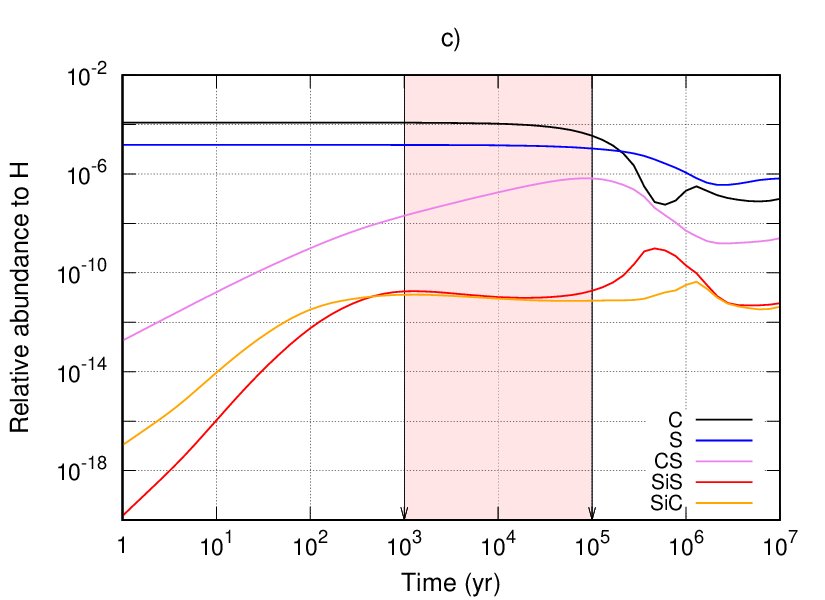}
\includegraphics[width=\columnwidth,keepaspectratio]{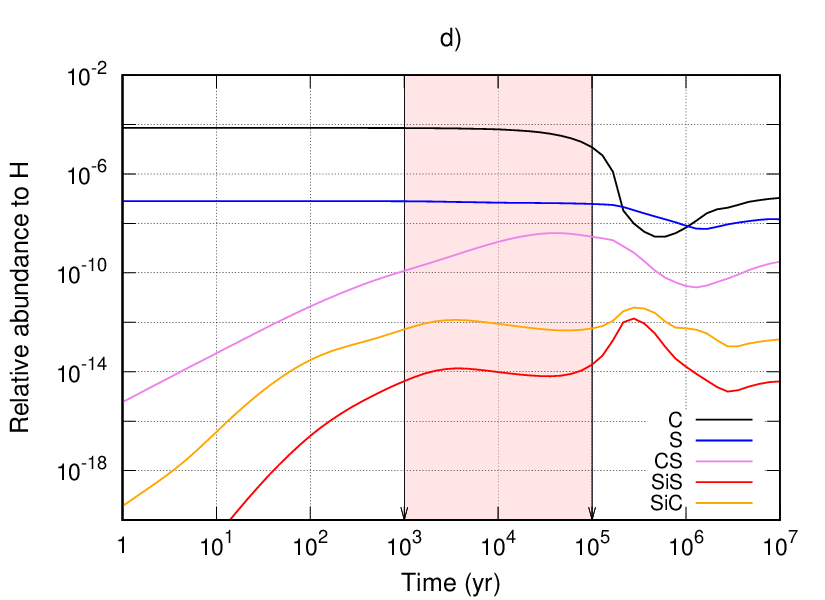}
\caption{\label{fig:models} Time evolution of the abundances of C, S, CS, SiS, and SiC, shown in black, blue, magenta, red, and yellow colours, respectively. Panels a) and b) illustrate simulations with high and low metallicity values, respectively, using the default network of chemical reactions. Panels c) and d) represent high and low metallicity models, respectively,  incorporating the newly computed chemical reactions in this work (see Table~\ref{tab:net}). The range indicated by the arrows, between 10$^3$--10$^5$ yr, signifies an early stage of relative stabilization in the SiS abundance (see \S~\ref{sec4.1} ).}
\end{center}
\end{figure*}

The results are presented in the panels of Fig~\ref{fig:models}, displaying the predicted abundances of C, S,  CS, SiS and SiC over a 10-million-year duration in a molecular cloud. These panels emphasise two fundamental aspects. Firstly, they  contrast the impact of two distinct environments, one with high and the other with low metallicity, on molecular abundances. Secondly, they  compare
the chemical abundances computed using the KIDA default list of chemical reactions with those obtained after integrating the newly studied reactions into the dataset, as indicated in  Table~\ref{tab:net} as \lq\lq this work.\rq\rq

The high-metallicity abundance models significantly promote the production of SiS. Early on, the formation of SiS occurs, resulting in abundances that can be nearly four orders of magnitude higher than those obtained in the low-metallicity model. Figs~\ref{fig:models}a and \ref{fig:models}b depict models using the standard network of chemical reactions to simulate molecular clouds with high and low-metallicity values, respectively. In these two scenarios, the mechanisms responsible for the destruction of SiS become relevant around an age of $t=10^3$~yr. At this specific point in time, SiS is predominantly destroyed ($\sim$~100~\%) through ion-molecule reactions involving the cations S$^+$, He$^+$, and C$^+$. In contrast, Figs.~\ref{fig:models}c and \ref{fig:models}d exhibit the results when   
the first six chemical reactions listed in Table~\ref{tab:net} are also considered. These figures illustrate the estimated abundances for high and low metallicity values, respectively. When examining these results, it becomes evident that the first six chemical reactions significantly influence the SiS abundance, especially in high-metallicity environments.  The percentage contributions of formation and destruction reactions fluctuate over time, affecting the calculated abundances. Around the age of $10^3$~yr, a point during which SiS abundances partially stabilise, we note that the newly incorporated SiS+C channel emerges as the predominant mechanism of destruction, making a substantial contribution to $\sim$~98\% and 93\% of the destruction processes in the high and low metallicity models, respectively. The remaining percentage is attributed to ion-molecule reactions involving S$^+$, He$^+$, and C$^+$.

In regard to the steady state condition, the models in Fig.~\ref{fig:models} do not evolve unequivocally to such a state (e.g. \citealt{Lee1998}). Between 10$^5$--10$^6$~yr, models exhibit an abundance ripple. In early-time chemistry, between 10$^3$ and 10$^5$~yr, the high and low metallicity models attain in general a quasi steady state condition. In addition, the low and high metallicity models seems to evolve to a late steady state condition after 10$^6$~yr. Follow-up studies are needed to investigate deeper into the oscillations and the variable kinetics concerning the chemistry of Si- and S-bearing molecules. The steady state condition of these molecules may depend on various factors such as initial abundances, computed chemistry, and physical parameters like temperature and cosmic ray ionization rate (e.g. \citealt{Roueff2020,Dufour2023}).

It is important to highlight that the incorporation of the new neutral-neutral reactions presented in this work affect the abundance of SiS.  These new reactions accelerate the formation of SiS, making it appear at much earlier ages,  {$t\ll 10^3$~yr}, especially in scenarios characterised by high metallicity.  For the time $t \approx 10^3-10^5$~yr, the significance of the new destruction reactions is evident by around one order of magnitude of the relative abundance when comparing Figs.~\ref{fig:models}a and \ref{fig:models}c and Figs.~\ref{fig:models}b and \ref{fig:models}d for both scenarios, the high and low metallicity. Furthermore, despite the introduction of the SiS+C channel as an efficient mechanism for its depletion, the overall effect leads to an increase in the SiS abundance at the model's final time around  ($t=10^6-10^{7}$~yr) by about one and two order of magnitude for the high and low metallicity, respectively (see displays in Figs.~\ref{fig:models}). This increase is attributed to its formation through the SiC+S reaction, which we have computed in this work.

\subsection{Full chemical network}
\label{sec:4.2}

Open questions persist about molecular abundances in interstellar clouds, particularly as more chemical reactions are included in our analysis. Here, we present a model for a molecular cloud encompassing all reactions in Table~\ref{tab:net}, expected to shed light on the complex chemical dynamics in environments characterised by both high and low initial chemical abundances.

\begin{figure}
\begin{center}
\includegraphics[width=\columnwidth,keepaspectratio]{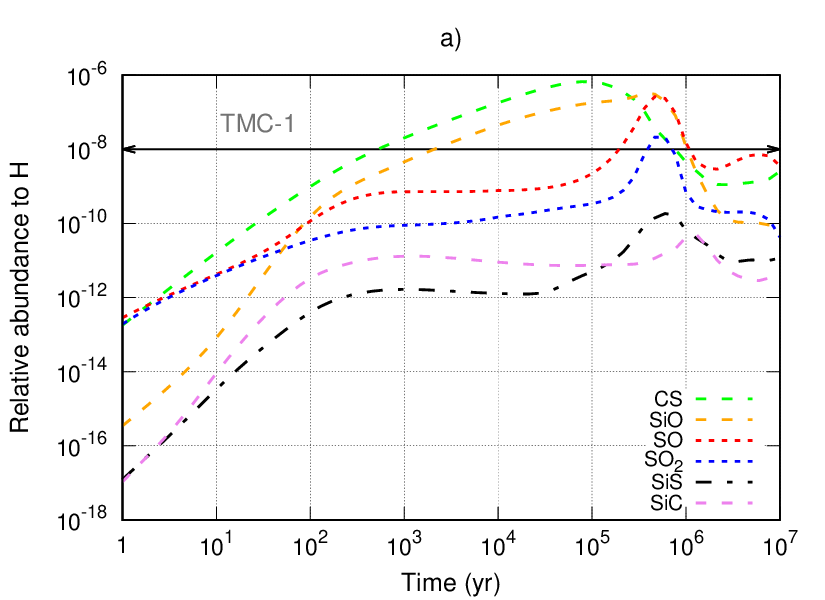}
\includegraphics[width=\columnwidth,keepaspectratio]{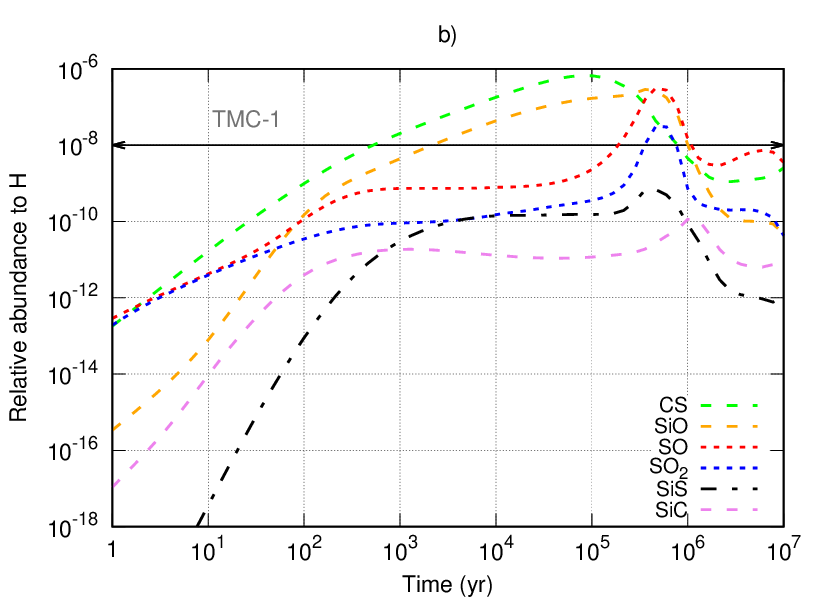}
\caption{\label{fig:models-b} Time evolution of the abundances of CS, SiO, SO, SO$_2$, SiS and SiC under the interstellar conditions $T=$10~K, $n_{\text H}$=2$\times$10$^4$ cm$^{-3}$.  a) Accounting for all the newly computed chemical reactions in this work listed in Table~\ref{tab:net} as well as the KIDA default network, and b) only employing the KIDA default network of chemical reactions. The horizontal arrows illustrates the observed abundance of SO in TMC-1  \citep{Lique2006} for comparison.}
\end{center}
\end{figure}

Table~\ref{tab:net} compiles various chemical reactions involving SiO, SO, SO$_2$, SiS, SiC, and other molecules, including recent reactions from this study and that of prior publications, enhancing our contemporary analysis of chemical processes. To assess the effects of these six initial reactions (Figs.~\ref{fig:models}), we conducted computational models in two distinct environments with varying elemental abundances, one low in metallicity and the other high. However, by running an extended model that incorporates all reactions listed in Table~\ref{tab:net}, we were able to analyse new pathways for SiS-related species. The simulation adopted the same physical conditions as described in (\S~\ref{sec4.1}), with gas temperature at 10 K and H$_2$ density at around 2 $\times$ 10$^4$ cm$^{-3}$. This model employed initial abundances based on high metallicity values and integrated the chemical reactions from Table~\ref{tab:net} into Nautilus' default chemical network.  Fig.~\ref{fig:models-b}a illustrates the results of the expanded model and it is compared with Fig.~\ref{fig:models-b}b which exhibits the results of the model using the default chemical network. These results depict the time-dependent abundances of CS, SiO, SO, SO$_2$, SiS, and SiC over a timescale of 10$^7$ yr. To illustrate the significance of reactions within the extended model, we have compiled the primary formation and destruction reactions in Table~\ref{table-reactions} as percentages at a time of 10$^3$~years.

A particularly notable distinction of this extended model is its contrast with the default KIDA network. In the default network, SiS formation primarily relies on the dissociative recombination reaction:
\setcounter{equation}{12}

\begin{align}\label{Eq2}
\text{HSiS}^+ + \text{e}^- \longrightarrow \text{H} + \text{SiS.}
\end{align}

\noindent It is worth noting that this mechanism can also lead to the formation of different species, which may compete with SiS formation:

\begin{align}
\text{HSiS}^+ + \text{e}^- \longrightarrow \text{Si} + \text{HS.}
\end{align}

\noindent This route contributes to the synthesis of Si and HS and, therefore, influencing the abundance of SiS. Considering the significance of HSiS$^+$, it is noteworthy that its formation can occur through the ion-neutral reactions:

\begin{align}
&\text{H}_2 + \text{SiS}^+ \longrightarrow \text{H} + \text{HSiS}^+, \label{Eq4}\\
&\text{SiS} + \text{HCO}^+ \longrightarrow \text{CO} + 
\text{HSiS}^+,\\
&\text{SiS} + \text{H}_3^+ \longrightarrow \text{H}_2 + \text{HSiS}^+.
\end{align}

\noindent Among these reactions, the  Eq.~\ref{Eq4} is the most important one, as it involves the reactant SiS$^+$, from which its primary formation routes are: 

\begin{align}
&\text{SiH} + \text{S}^+ \longrightarrow \text{H} + \text{SiS}^+,\\
&\text{OCS} + \text{Si}^+ \longrightarrow \text{CO} + 
\text{SiS}^+,\\
&\text{SiS} + \text{S}^+ \longrightarrow \text{S} + \text{SiS}^+.
\end{align}

In the extended model, within the ambit of the five most important SiS formation pathways (as shown in Table~\ref{table-reactions}), we find that the KIDA default mechanism presented in  Eq.~\ref{Eq2} is positioned fourth in terms of significance, based on its percentage contribution. The most significant pathways in Table~\ref{table-reactions} are associated with neutral-neutral reactions, including a novel pathway introduced in this study. Based on these findings, we have demonstrated the importance of studying neutral-neutral processes in the formation of interstellar molecules. The reactions involving Si with SO, HS, and H$_2$S have the potential for contributing to the production of SiS under interstellar conditions. Additionally, our aim has been to conduct a comprehensive analysis of the S+SiC system, which produces C and SiS as byproducts. In the context of astronomical observations, it is worth noting that while these molecular species have been reported sporadically,  SiC \citep{Cernicharo1989} and SiS  \citep{VP2019} have been definitively detected within the circumstellar envelope of IRC+10216 —a source known
for its extensive carbon chemistry. In the protostellar shock region L1157-B1, the detection and formation of SiS have also been investigated \citep{POD17:L16,MOT21:37}. Further investigations will be carried out to explore the presence of SiC in other sources.

Neutral-neutral reactions play a pivotal role in reducing the gas abundance of SiS. Specifically, when we consider the reference age of $t=$1$\times$10$^3$~yr, the reaction between SiS and O, leading to the formation of SiO and S, is particularly noteworthy with a
percentage contribution of 95.9~\% (Table~\ref{table-reactions}). It stands out as the most significant mechanism. This reaction has been identified as a crucial mechanism responsible for the depletion of SiS in the envelopes surrounding evolved stars~\citep{ZAN18:38}, as well as in models representing protostellar shock regions~\citep{MOT21:37}. Additionally, the reaction between SiS and C, resulting in Si and CS, stands as the second most important mechanism, however with a low contribution of only 4.0~\%. In addition to these two primary neutral-neutral mechanisms, ion-neutral reactions also have a role, although they are less prominent in this context. This is exemplified by the SiS+S$^+$ reaction (Table~\ref{table-reactions}).

\begin{table}
\caption{Contribution of the primary formation and destruction reactions of SiS at $t$=1$\times$10$^3$~yr obtained from  the model using the full chemical network, i.e., KIDA reactions plus the new ones studied in this work. The model considers a molecular cloud with $T$=10 K, $n$(H$_2$)=2$\times$10$^4$ cm$^{-3}$ and a cosmic ray ionisation rate of 1.3$\times$10$^{-17}$~s$^{-1}$  (see Fig.\ref{fig:models-b}a)}.
	\label{table-reactions}
    \centering
	\begin{tabular}{ll}
		\hline
		\textbf{Reaction} &  Contribution [\%] \\
\hline
\it{Formation of SiS} & \\
Si + HS        $\longrightarrow$ H     + SiS & 30.6 \\
S + SiC        $\longrightarrow$ C     + SiS & 26.5 \\
Si + SO        $\longrightarrow$ O     + SiS & 23.8 \\
HSiS$^+$ + e$^- \longrightarrow$ H     + SiS & 18.3 \\
Si + H$_2$S    $\longrightarrow$ H$_2$ + SiS & 0.70 \\
  \hline
\it{Destruction of SiS} &  \\
SiS + O         $\longrightarrow$  SiO   + S        & 95.9 \\
SiS + C         $\longrightarrow$  Si    + CS       & 4.00 \\
SiS + S$^+$     $\longrightarrow$  S     + SiS$^+$  & 0.10 \\
\hline
\end{tabular}
\end{table}

In Fig.~\ref{fig:models-b}a and Fig.~\ref{fig:models-b}b,  in addition to the SiS result, we have incorporated the abundance curves of CS, SiO, SO, SO$_2$, and SiC. This inclusion serves two primary purposes. Firstly, these chemical species collectively form a chemical network involving Si- and S-bearing molecules that ultimately leads to the formation of SiS. Secondly, these species have been observed through vibrational and rotational spectral lines, enabling us to discuss the computed abundances. In a comprehensive study involving 199 molecular clumps, \citet{Li2019} identified SiO 5-4 emission in 102 of these sources, finding relative abundances concerning molecular hydrogen ranging from 1.1 $\times$ 10$^{-12}$ to 2.6 $\times$ 10$^{-10}$. Their analysis revealed that SiO abundance does not diminish as massive star formation advances within these molecular clumps. From Local Thermodynamic Equilibrium (LTE) analyses, they assumed excitation temperatures of 18~K for Infrared Dark Clouds (IRDCs), 25~K for protostars, and 30~K for H II regions. Compared to the extended model, the predicted abundance of SiO exhibits a variation, ranging from 10$^{-16}$ to 10$^{-8}$. Across most of the model, these predicted abundances surpass the values estimated by \citet{Li2019}. However, considering the significant variability that can arise in SiO abundances, e.g., across different massive star forming regions, which can differ by orders of magnitude \citep{Guerra-Varas2023}, the comparison and correlation between observational and computed molecular abundances warrant further investigations.

Examining the molecule SO within TMC-1, \citet{Lique2006} conducted an analysis that unveiled abundance values around 2.5 $\times$ 10$^{-8}$. This high abundance was observed in a source structure consisting of multiple cores with $T_{\emph{kin}}$=8~K and $n$(H$_2$) $\sim$ 3 $\times$ 10$^4$ cm$^{-3}$ surrounded by an envelope with $T_{\emph{kin}}$=10~K and $n$(H$_2$) $\sim$ 6--8 $\times$ 10$^3$ cm$^{-3}$. 
In a study of sulphur-bearing molecules in the Horsehead Nebula, \cite{Riviere2019} identified
13 S-bearing species across two distinct targets: the photodissociation region and the core of this source. On the core they estimated low excitation temperature values of $\sim$ 8.2 and 6.9 K for SO and SO$_2$, respectively, as well as molecular abundances of $\sim$ 8.6$\times 10^{-10}$ and 5$\times 10^{-11}$, respectively. Consistent with the observations, the simulation in this  cold environment not only predicts abundances within the same order of magnitude, but also indicates that SO is produced more efficiently than SO$_2$. Consequently, SO molecule, in comparison with SO$_2$ exhibits a higher abundance throughout the course of time evolution, mainly after 10$^2$ yr (see Fig.~\ref{fig:models-b}). In a recent study, \cite{Law2023} identified the compact SO emission within the HD 169142 protoplanetary disc. They also reported the first tentative detection of SiS. In this context, they suggest that the abrupt rise in temperature and density within the shocked gas surrounding the embedded planet likely leads to a hot gas-phase chemistry.  \citet{Law2023} discussed the efficiency of the reaction Si+SH$\longrightarrow$SiS + H under warm temperatures (200~K), which is similar to those observed in the gaseous environment surrounding an embedded giant planet. Similarly, they discussed the Si+SO $\longrightarrow$SiS + O route considering the SO detection in HD16142.

In addition to the investigation of the S+SiC reaction conducted in this study, we have identified three additional reactions that can contribute to the formation of SiS. These newly revised reactions involve Si+HS, Si+SO and Si+H$_2$S, as detailed in  Table~\ref{table-reactions}. Therefore, our findings enhance the comprehension of neutral-neutral reactions, which, when compared to ion-neutral mechanisms, can also play a significant role in impacting molecular abundances and the chemistry of the ISM.

\section{Conclusions and Perspectives}
\label{sec:conc}

In this work, we have performed accurate electronic structure calculations, including explicitly correlated wave functions, to elucidate new destruction and formation paths for the SiS molecule. We investigated the energetic viability of these reactions and estimated their rate coefficients using statistical theories. The results indicate that reactions between SiC and atomic sulphur are barrierless and can produce both SiS and CS at similar rates. Moreover,  the destruction of SiS by atomic carbon does not require an activation energy and has even larger rate coefficients. These reactions can occur rapidly, even at the very low temperatures in the ISM.

The temperature-dependent rate coefficients  for reactions involving Si- and S-bearing molecules, investigated in this work, were fit to the expression $k(T)=\alpha \left( T/300 \right)^{\beta} \exp{(-\gamma/T)}$. These rate coefficients can be incorporated into astrochemical databases, alongside previously published computational and experimental findings from the literature, all of which are summarised here.

Astrochemical gas models were conducted to assess the impact of previously uninvestigated reactions on the relative abundances of C, S,  CS, SiS and SiC. These simulations incorporated two distinct sets of initial chemical abundances characteristic of low- and high-metallicity environments. While most astrochemical studies focus on high metal abundances, our research highlights the significance of low-metal models, which are applicable to both the galactic halo and extragalactic sources. The computations for these models were performed under conditions of gas temperature $T$=10~K and density $n_{\text H}$=2$\times$10$^4$ cm$^{-3}$, visual extinction of $A_V$=10~mag, and a cosmic rays ionisation rate of  $\zeta$=1.3$\times$10$^{-17}$ s$^{-1}$. Molecular abundances were computed over a timescale of  10~Myr.

The high-metallicity abundance models significantly enhance the production of SiS, even during its early formation, resulting in abundances nearly four orders of magnitude greater than those observed in the low-metallicity counterpart. This enhancement can be attributed, in part, to the relatively large
initial abundances of C, S, and Si, which are higher by factors of $\sim$~2, 190, and 210, respectively, compared to the low-metallicity conditions. In both models, a threshold of 10$^3$~yr emerges as the point at which destruction mechanisms begin to impact the SiS production. Notably, two reactions, C + SiS $\longrightarrow$ Si+CS and S + SiC $\longrightarrow$ C + SiS, which were computed in this work, emerge as significant mechanisms for estimating the abundance of SiS.

We extended the KIDA database with previously studied reactions and newly computed ones.
This extended chemical network was then utilised to determine the abundances of the molecular species SiO, CS, SO, SO$_2$, SiS, and SiC, as they have been previously detected as well as linked to the SiS chemistry.
Through this approach, we simulated a model for a molecular cloud with initial abundances typical of a high-metallicity environment. This extended model has contributed to the discovery of new formation pathways for SiS.

In addition to the HSiS$^+$ + e$^- \longrightarrow$ H + SiS route, which was the sole pathway recognised in KIDA, this work has introduced alternative channels, such as Si+HS and S+SiC. These two revised and newly computed routes serve as alternatives for SiS formation. This study also analysed the destruction channels of SiS.
The rapid interaction between SiS and neutral elements, such as O and C, substantially diminishes the abundance of SiS. 
These reactions involving neutral elements outweigh those involving ions or molecular ions.
In conclusion, this study 
emphasises the crucial role played by neutrals and radicals in the chemical processes occurring  within the cold and dense interstellar regions.

Apart from SiS, our study also provided time evolution curves for the abundances of CS, SiO, SO, SO$_2$, and SiC. Their predicted abundances were discussed considering observational values, particularly for the relevant species SiO and SO. Our simulations confirm the relatively high abundance and efficient production of SO when compared to SO$_2$ in cold environments. In the context of protoplanetary discs, the potential significance of SiS formation pathways, such as Si+HS, Si+SO, and S+SiC becomes evident.
Our research enables us to gain insights into the reaction dynamics of Si- and S-bearing molecules, estimate their abundances, and sheds light on the formation and destruction pathways associated with the SiS molecule. From a broader perspective, a further research is necessary to tackle the intriguing questions regarding the steady state condition, the influence of the kinetic on it, and the percentage contribution of the different chemical species over time.

\begin{acknowledgements}

We would like to acknowledge the anonymous reviewer for constructive and insightful comments. The authors would like to thank the financial support provided  by the Coordena\c c\~ao de Aperfei\c coamento de Pessoal de N\'ivel Superior - Brasil (CAPES) - Finance Code 001,
Conselho Nacional de Desenvolvimento Cient\'ifico e Tecnol\'ogico (CNPq), grants 311508/2021-9 and 405524/2021-8, Funda\c c\~ao de Amparo \`a Pesquisa do estado de Minas Gerais (FAPEMIG), and Centro Federal de Educação Tecnológica de Minas Gerais (CEFET-MG). Rede Mineira de Química (RQ-MG) is also acknowledged. E.M. acknowledges support under the grant "María Zambrano" from the UHU funded by the Spanish Ministry of Universities and the "European Union NextGenerationEU." This project has also received funding from the European Union's Horizon 2020 research and innovation program under Marie Sklodowska-Curie grant agreement No. 872081, grant PID2019-104002GB-C21 funded by MCIN/AEI/10.13039/501100011033, and, as appropriate, by "ERDF A way of making Europe," the "European Union," or the "European Union NextGenerationEU/PRTR." This work is also supported by the Consejería de Transformación Económica, Industria, Conocimiento y Universidades, Junta de Andalucía and European Regional Development Fund (ERDF 2014-2020) PY2000764.
The authors also acknowledge the National Laboratory for Scientific Computing (LNCC/MCTI, Brazil) for providing HPC resources of the SDumont supercomputer, which have contributed to the research results reported within this paper. URL: http://sdumont.lncc.br
\end{acknowledgements}

%
%

   \bibliographystyle{aa} 
   \bibliography{varbre,SiS} 
\end{document}